\documentclass[conference]{IEEEtran}
\usepackage{biblatex}
\usepackage{amsmath,amssymb,amsfonts}
\usepackage{algorithmic}
\usepackage{graphicx}
\usepackage{textcomp}
\usepackage{xcolor}
\usepackage{fancyhdr}
\usepackage{bigstrut}
\usepackage{multicol}
\usepackage{multirow}
\usepackage{colortbl}
\usepackage{float}
\usepackage{hyperref}
\usepackage{url}
\usepackage{setspace}
\def\BibTeX{{\rm B\kern-.05em{\sc i\kern-.025em b}\kern-.08em
    T\kern-.1667em\lower.7ex\hbox{E}\kern-.125emX}}
    
\begin{document}
 
\title{Towards a Cloud-Based Ontology for Service Model Security - Technical Report}

\author{
  Mohammed Kharma $^*$\\
  \textit{Birzeit University}\\
  \textit{Ramallah, Palestine}\\
  \textit{mkharmah@birzeit.edu}
 
  \and
  Ahmed Sabbah \\
 \textit{Birzeit University}\\
 \textit{Ramallah, Palestine}\\
 \textit{asabah@birzeit.edu}\\

 \and Mustafa Jarrar \\
 \textit{Birzeit University}\\
 \textit{Ramallah, Palestine}\\
 \textit{mjarrar@birzeit.edu}

 }



\maketitle
\thispagestyle{fancy}

\begin{abstract}
The adoption of cloud computing has brought significant advancements in the operational models of businesses. However, this shift also brings new security challenges by expanding the attack surface. The offered services in cloud computing have various service models. Each cloud service model has a defined responsibility divided based on the stack layers between the service user and their cloud provider. Regardless of its service model, each service is constructed from sub-components and services running on the underlying layers. In this paper, we aim to enable more transparency and visibility by designing an ontology that links the provider's services
with the sub-components used to deliver the service. Such breakdown for each cloud service sub-components enables the end user to track the vulnerabilities on the service level or one of its sub-components. Such information can result in a better understanding and management of reported vulnerabilities on the sub-components level and their impact on the offered services by the cloud provider. Our ontology and source code are published as an open-source and accessible via GitHub: \href{https://github.com/mohkharma/cc-ontology}{mohkharma/cc-ontology} 
 
\vspace{0.5cm}
\textit{Keywords} — Cloud Computing Ontology, Cloud Computing, Security, CVE.
\end{abstract}

\section{Introduction}
\label{sec:intro}

Vulnerability analysis is a technique for identifying and evaluating potential software applications or system security hazards \cite{DBLP:journals/ress/Aven07}. It involves identifying the various threads or methods that an attacker could use to compromise the security of the system and then analyzing the security dangers associated with each thread. Vulnerability analysis is important to assist software developers, security professionals, and end users in understanding a system's attack surface and identifying potential vulnerabilities. By analyzing and identifying the system's vulnerabilities while contemplating the perspective of the attacker. To enhance safety, greater community-based efforts are required to share information about them \cite{hemberg2020linking}. 

There is a considerable increase in the number of security vulnerabilities, the NVD (National vulnerability database) reported new 8,051  vulnerabilities in the first quarter of 2022 with an approximate 25\% increase over the same quarter of the previous year \cite{25CyberS64:online}. On the research side, the security domain is prioritized with the objective of limiting or totally mitigating the impact of different security threats and attacks. Most of the previous studies focused on building a knowledge model to describe security information such as Common Vulnerabilities and Exposures (CVE), Common Weakness Enumeration (CWE), and Common Attack Pattern Enumeration and Classification (CAPEC) \cite{Brazhuk2019SemanticMO, Brazhuk2020framework,hemberg2020linking, KharmaTaweel22}. Different methods exist for representing such security knowledge. A method proposed by Brazhuk et. al. is based on semantic (ontology) modeling \cite{Brazhuk2019SemanticMO, Brazhuk2020framework}. To define the terms and relationships between various security information that can be used to integrate them. The graph-based method presented in \cite{hemberg2020linking} utilized the graph presentation to store and consolidate a comprehensive set of security information while maintaining the relationships between various security information records. The literature focused on linking the vulnerability information with the associated products as the NVD database provides the affected products and versions as part of the vulnerability report  \cite{hemberg2020linking}. 

Our strategy in this paper is to link the cloud provider services with the sub-components that the cloud provider uses to provide a particular service. Such breakdown for each cloud service sub-components enables the end user to track the vulnerabilities on the service level or one of its sub-components. For example, a cloud provider offers Serverless services, where vulnerabilities at the runtime layer, such as the Java Runtime Environment (JVM), can have
implications for the end users of Serverless service. So, tracking the vulnerabilities on the service (Serverless)
and sub-component (JVM, or other sub-components) levels can help the end user to be aware of potential vulnerabilities that can affect his usage of the Serverless service. Therefore, we ensure a holistic understanding of potential risks, enabling better
protection and proactive measures to address vulnerabilities in the cloud environment. Such an approach can push for more transparency between the cloud provider and the end users.

The rest of this paper is organized as follows, section \ref{sec:relatedWork} presents recent research on ontology-driven threat modeling. Section \ref{sec:methodology} describes our approach to link the
provider’s services with the sub-components used
to deliver the service. Section \ref{sec:Evaluation} discusses the evaluation of the created ontologies. Finally, section \ref{sec:conclusion} concludes the work and provides future directions for our research.

\section{Related work}
\label{sec:relatedWork}

According to our previous study, we provided a narrative literature review for threat modeling approaches in cloud computing \cite{KharmaTaweel22}. It was found that the Modelling language-based approach driven by ontologies was a well-organized approach that provided support for automatic threat modeling.
Ontologies are used to represent agreed domain semantics \cite{jarrar2009ontology} and to enable the reusability of such semantics \cite{JM02b}. Ontology-Driven Threat Modelling (OdTM) framework, as described by the OWASP foundation in \cite{OWASPOnt78:online}, is one of the most recent contributions to ontology-driven threat modeling. This study extends their ontological work through and connects the cloud computing service models with the Common Vulnerabilities and Exposures (CVE). A relevant research is described in Brazhuk and Olizarovich's published works  \cite{Brazhuk2020framework,Brazhuk2021Towards}. The OdTM framework purpose is to automate the process of threat modeling to a certain degree by utilizing ontology to compile all information pertaining to the security of a specific domain. It relies on building domain-specific threat modeling-based ontology on top of the base threat model ontology to address the needs of specific domains by extending the base model with domain-specific ontologies. 

There are numerous publicly available data sources that can aid in building a knowledge base and cloud ontology that target the security aspect. Brazhuk ~\cite{Brazhuk2019SemanticMO, Brazhuk2021Towards} suggested an OWL(Web Ontology Language) ontology based on the Enterprise Matrix of MITRE ATT\&CK, CAPEC (Common Attack Pattern Enumeration and Classification),  CWE (Common Weakness Enumeration) and partial CVE (Common Vulnerabilities and Exposures) data, and it can answer questions about grouping and classification of security concepts based on supplied criteria using DL (Description Logic) and RDF query language (SPARQL) queries. Brazhuk concluded that existing security enumerations have weak links both quantitatively and qualitatively. To get the maximum benefits out of these data sources and at least from the CVE, we aimed to include all CVEs either as it has a relation with CWE or is linked with CPE. Another relevant effort carried out by Massachusetts Institute of Technology researchers ~\cite{hemberg2020linking}, they proposed BRON which is a relationship graph in which the entries of its many information sources are represented as certain sorts of vertices and their internal as well as external links as edges. BRON utilized all mentioned data sources used by Brazhuk, but as a graph instead of a semantic model including additional CVE information from NIST including "severity score" and "Known Affected Hardware or Software Configuration" fields.

On the other hand, Valja et. al~\cite{valja2020automating} tried to provide information on how to utilize existing data and incorporate domain knowledge through ontologies to enhance the system's security.  The role of ontology in their paper is to enrich the data obtained from different types of data sources by applying various functions. The ontology framework functions are invoked from the modeling automation process adapters, which process files from specific data sources. The goal of the ontology framework is to enhance the data by adding context and knowledge. This enriched data is then used as input for the modeling automation process. The ontology helps in standardizing and classifying the data, improving the semantic accuracy of the threat models. The authors claimed that ontology plays a crucial role in improving the quality and precision of automated threat models.

Moreover, another research \cite{Aranovich_2022} discussed the use of ontologies in the Semantic Web to enhance cybersecurity measures. It describes the development of an ontology called TRONTO, which extends the information in the National Vulnerability Database (NVD) by inferring new classes, enriching relations, and expanding conceptual coverage. The ontology is used to search for and query social media threads that contain cybersecurity-related information, and natural language processing techniques are used to relate unstructured information to concepts in the ontology. The paper highlights the advantages of Semantic Web technologies in integrating information from multiple and often heterogeneous sources, without human intervention. Rosa et.al.~\cite{Rosa2022} presented a novel ontology-based approach to utilize ontology to identify and map threats to assets. With the support of formally sound approaches, this process can be streamlined and made more efficient. From an ontology perspective, the authors introduced ThreMA, an ontology-based approach for automating threat modeling in ICT infrastructures. ThreMA provides a standard meta-model that describes the infrastructure and a set of rules for threat modeling. The meta-model consists of three ontologies modules: ICT ontology for modeling the infrastructure, Data Flow ontology for representing data flow diagrams, and threat ontology for characterizing threats. The use of ontology and inference rules allows for a syntactical representation of the problem, mimicking expert thinking. This approach enhances extensibility, maintainability, and integration in a rapidly changing context. The paper emphasizes the importance of using ontologies to address the lack of context and low accuracy in threat modeling. Overall, ThreMA offers a comprehensive ontology-based solution for automating threat modeling in ICT infrastructures.

\section{Methodology}
\label{sec:methodology}

This work presents an ontology for representing various data sources about cloud computing and security. This ontology enables a knowledge presentation framework for all cloud computing and its relationships. The ontology consists of several modules: Cloud Computing and services, Cloud Service underlying components, and CVE module. 

\subsection{Cloud Computing Stack and Services Ontology Module}
\label{sec:methodology_CC_Service}
\begin{figure}[ht]
\centering\includegraphics[scale=0.44]{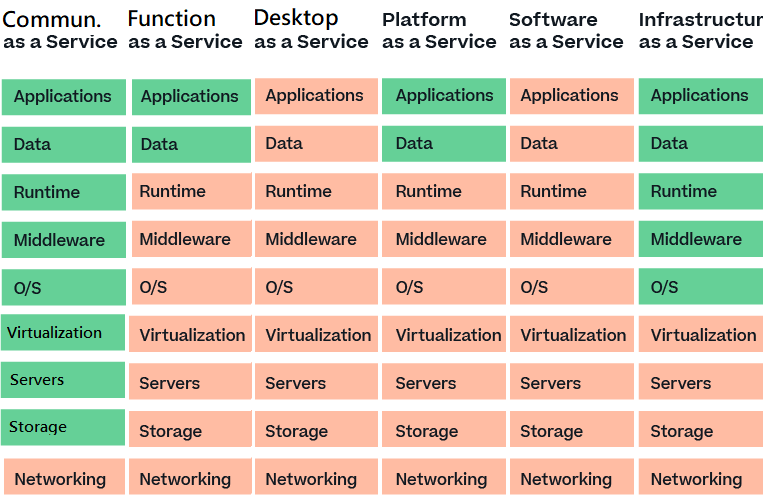}
\caption{Cloud computing stack. The green color depicts the layers managed by the used user. While the orange-colored boxes represent the layers managed by the cloud provider.}\label{cloudServiceModel}
\end{figure}

\begin{figure}[ht]
\centering\includegraphics[scale=0.39]{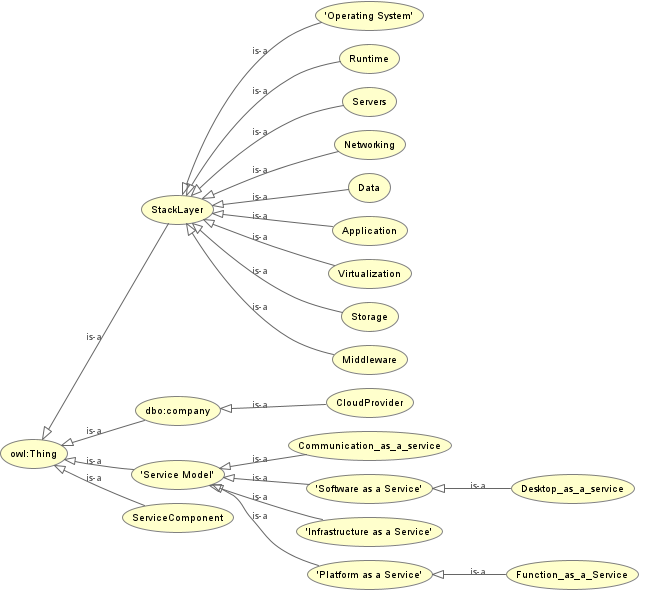}
\caption{Cloud computing stack ontology.}\label{CloudServiceModelOntology}
\end{figure}

This section represents our proposed ontology module that covers the cloud computing stack and services. 
Our extension ontology is provided as a separate ontology, which is an important design criteria in ontology engine engineering \cite{J05a, J05}.
This ontology can be used to unify and provide a primary baseline for cloud computing stack, threat understanding, and system diagram presentations. Firstly, we start by creating the ontology of the cloud computing stack. Figure \ref{CloudServiceModelOntology} depicts the ontology of the three cloud stack: Software as a Service (SaaS), Platform as a Service (PaaS), Infrastructure as a Service (IaaS), Function as a Service (FaaS), Communication as a service (CaaS), and Desktop as a service (DaaS). Figure \ref{cloudServiceModel} depicts the cloud service model. Cloud service has nine layers, green colored layers mean these layers are managed by the client while the other color refers to layers managed by the cloud providers.

The following is the list of classes that we created to describe the cloud service model, except the highlighted classes in italics. We adopted these \textit{classes} from previous work \cite{Brazhuk2019SemanticMO, Brazhuk2020framework}:

\begin{itemize}
    \item \textbf{Stack layer: } In the design of a cloud computing system, a superclass called StackLayer has been created. This class serves as a foundation for organizing various layers of the cloud infrastructure. Several subclasses have been derived from StackLayer to represent different components of the cloud environment: \textit{Application, Data, Middleware, Networking, Operating System, Runtime, Servers, Storage, Virtualization}.
    
    \item \textbf{Cloud provider: } To represent the cloud provider companies, we created a new class called CloudProvider that extends the company class from DBPedia \footnote{https://dbpedia.org/data3/company.json}.
    
    \item \textbf{Cloud services:} To cover the variety of cloud service models, we created the following classes that extend the Service Model: Communication as a service, \textit{Infrastructure as a Service}, \textit{Platform as a Service}, Function as a Service, \textit{Software as a Service}, and Desktop as a Service.

     \item \textbf{Service component:} This represents the sub-components of which a cloud provider service consists. We use this class to link the affected products by a particular CVE against the sub-components of a cloud service provider.
     
\end{itemize}

We use the following object properties to create the relations between different classes as follows:
\begin{itemize}
    \item \textbf{offerServices: }  The relationship between a CloudProvider and the ServiceModel it offers.
\item \textbf{provides: } The relationship between a ServiceModel and the specific components or resources it provides. For example, IaaS provides Networking, Servers, Storage, and Virtualization.
\item \textbf{hasComponent: } The relationship between  different Service models and their related sub-components ServiceComponent.

\item \textbf{componentImpactedByCVE: } Represents the relationship between ServiceComponent and its related Common Vulnerabilities and Exposures (CVE).
\end{itemize}

To give more clarity, we list the following individuals as examples that can be generated based on the proposed ontology to identify the CVE for each component:
\begin{itemize}
    \item 
Amazon, Google, and Oracle: are individual cloud providers.
\item Microsoft Office 365, Azure AppService: These are individuals of services offered by cloud providers.
\item Azure RDP: This is an individual of applications provided as part of the cloud services.
\item  Azure Kafka, Azure Linux, Azure Java: All are individuals of middleware, operating systems, and runtime environments provided by cloud services.
\item Azure VLAN: A specific networking component provided by a cloud service.
\item Azure S3: A specific storage component provided by a cloud service.
\item Azure Hypervisor: A specific virtualization component provided by a cloud service.

\end{itemize}

\subsection{Cloud Service Underlying Components}
\label{sec:methodology_CC_Service_components}

In order to capture and establish connections between potential Common Vulnerabilities and Exposures (CVEs) that could impact a cloud service, we break down each service model into its separate sub-components. For example, Azure Kubernetes Service (AKS) from Microsoft has multiple components such as:

\begin{itemize}
    \item \textbf{Kubernetes: } AKS is based on the open-source Kubernetes project, which is an industry-standard container orchestration platform.

  \item \textbf{Docker: } Docker is a popular open-source platform that allows you to build, package, and distribute containerized applications. AKS utilizes Docker to manage and run containers on the worker nodes.

  \item \textbf{Containerd: } Containerd is an open-source container runtime that is used by AKS to manage container execution and lifecycle events. It provides a runtime environment for containers and implements the Open Container Initiative (OCI) specifications.
\end{itemize}

By adopting such a design approach, we can establish connections between CVEs not only at the cloud service level but also at the sub-component level. This enables the capture and highlighting of the potential impact of vulnerabilities on the proposed services by the cloud provider. For example, we can link specific CVEs, such as CVE-2021-24109 \footnote{https://nvd.nist.gov/vuln/detail/CVE-2021-24109} \footnote{https://vuldb.com/?id.169481}, directly to the cloud service itself which provides users with transparency and a comprehensive view of expected vulnerabilities. This taxonomy provides end users with insights into potential risks that might affect their services, enabling them to make informed decisions regarding security measures and risk mitigation.
The same approach can be applied to Serverless services, where vulnerabilities at the runtime layer, such as the Java Runtime Environment, can have implications for the end users of those services. By considering vulnerabilities at both the service and sub-component levels, we ensure an understanding of potential risks, enabling better protection and proactive measures to address vulnerabilities in the cloud environment

\subsection{CVE Data Integration Ontology Module}
\label{sec:methodology_CVE_Service}

CVE stands for Common Vulnerabilities and Exposures. It is a system used to identify and provide unique identifiers for publicly known information security vulnerabilities and exposures \cite{HomeCVE80:online}. The primary purpose of CVE is to provide a standardized and unique naming scheme for vulnerabilities and weaknesses found in software or hardware products.
Each CVE entry includes a specific identifier, such as "CVE-YYYY-NNNN," where YYYY represents the year the CVE was assigned, and NNNN is a sequential number. For example, CVE-2023-1234 would be the 1234th CVE entry assigned in the year 2023. When a security vulnerability is discovered in a software application, operating system, hardware device, or any other technology, the details of the vulnerability are reported to the CVE project. The project's team validates the vulnerability and assigns a unique CVE identifier to it \cite{HomeCVE80:online}.

\begin{figure}[ht]
\centering\includegraphics[scale=0.80]{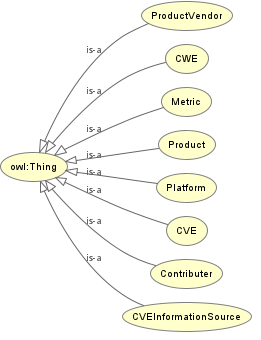}
\caption{Common Vulnerabilities and Exposures (CVE) ontology}\label{CVE_ontology}
\end{figure}

We created an ontology to represent the CVE information. Also, we created a parser that converts the unstructured security information represented by CVE into OWL format which is considered CVE ontology. Figure \ref{CVE_ontology} depicts the CVE ontology.

The parser pipeline consists of the following:

\begin{itemize}
    \item Download the CVE records from \cite{Download55CVE}.
    \item Parse the CVE record in JSON format.
    \item Use the JSON file to map with CVE ontology.
    \item Create a new RDF format file for the input CVE record.
    \item Link the CVEs to the cloud services components based on the affected product information provided in the CVE report.
\end{itemize}
The Java version of this parser can be found on Github \footnote{https://github.com/mohkharma/cc-ontology}

The list of classes that we use to describe the CVE data is as follows:

\begin{itemize}
\item \textbf{Common Vulnerabilities and Exposures (CVE): } Is the class we use it instantiates individuals from each CVE record and uses the CVE Id as the URI of the instance. For example, CVE-2021-24109 \footnote{https://nvd.nist.gov/vuln/detail/CVE-2021-24109} \footnote{https://vuldb.com/?id.169481} is an instance of Common Vulnerabilities and Exposures (CVE) class.
\item \textbf{CVEInformationSource: } This class captures vulnerability information related to a specific CVE ID provided by a specific organization participating in the CVE program.
\item \textbf{Contributor: } This represents the identity that takes credit for researching, discovering, remediating, or helping with activities related to this CVE.


\item \textbf{Metric:}  A class that provides the impact scores of the CVE. Each instance of this class contains the base score (e.g., 6.4). Linked with scoring system standard version (e.g., 3.1). And has values of matrix attributes Attack Vector, Attack Complexity, Privileges Required, User Interaction, Scope, Confidentiality, Integrity, and Availability. Values of these scoring attributes for example could be Physical, Low, None, None, Unchanged, High, High, and High, respectively.

\item \textbf{Platform: } This is the platform class that the particular vulnerability is applicable, such as Windows operating system.





\item \textbf{Product:} Which represents the product(s) that impacted by the reported CVE. We use this class to link the affected products by a particular CVE against the ServiceComponent concept in the cloud services ontology. Nginx \footnote{https://www.cvedetails.com/vulnerability-list/vendor\_id-10048/Nginx.html} is an instance of the Product class. 

\item \textbf{ProductVendor:} Which represents the vendor of the product(s) that are impacted by the reported CVE. We use this class to link the affected products by a particular CVE against the CloudProvider concept in the cloud services ontology. Microsoft and Google are examples of instances of ProductVendor class.

\item \textbf{Common Weakness Enumeration (CWE):} Is the class we use it instantiates individuals from each Common Weakness Enumeration (CWE) in order to link it with the CVE instances. For example, CWE-475 \footnote{https://cwe.mitre.org/data/definitions/475.html} is an instance of CWE class. 

\end{itemize}

\begin{figure}
    \centering
    \includegraphics[scale=0.75]{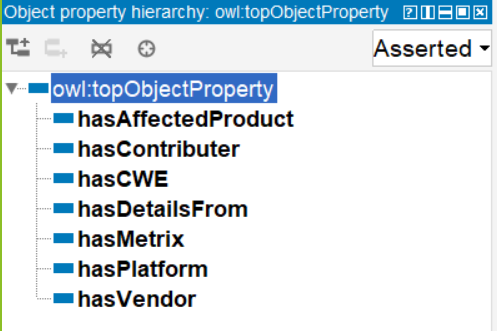}
    \caption{CVE ontology object proprieties}
    \label{fig:CVE_properities}
\end{figure}

\begin{figure}
    \centering
    \includegraphics[scale=0.55]{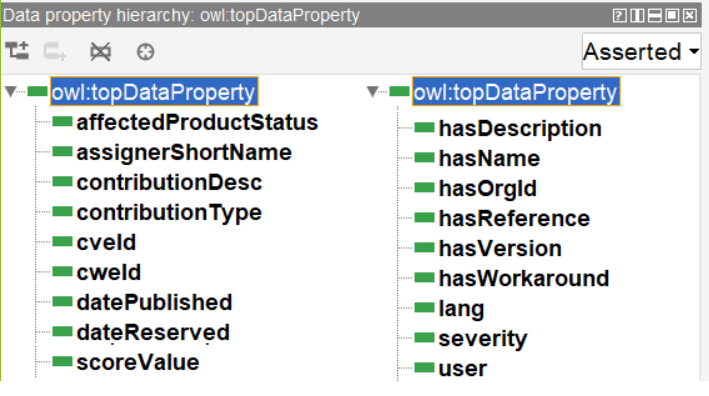}
    \caption{CVE ontology data type proprieties}
    \label{fig:CVE_datatype_properities}
\end{figure}

We use the object properties to create the relations between different concepts. Figure \ref{fig:CVE_properities} depicts the 16 object properties created in the CVE ontology. Also, we created 27 data type properties as shown in figure \ref{fig:CVE_datatype_properities} depicts the 16 object properties created in the CVE ontology.

\vspace{.15cm}

\section{Building a cloud services knowledge graph }
\label{sec:Evaluation}
In order to evaluate the effectiveness of the proposed cloud computing stack and CVE ontology modules, we developed automation tools utilizing the Apache Jena Java and Jackson libraries \cite{ApacheJe10:online} \cite{GitHubgo79:online}. We use these tools in order to extract the CVEs records and save them in RDF format. By leveraging the structure and concepts defined in the proposed ontologies, we constructed a comprehensive knowledge graph that encompasses these individuals. Such a knowledge graph serves as a valuable resource for analyzing and understanding the cloud computing service's implicit vulnerabilities by drawing the focus on the sub-component vulnerabilities that might affect the offer of cloud service. In addition, to retrieve information from the generated knowledge graph, we used SPARQL to create queries to extract specific information and explore the relationships between entities. This information presents the relationships between the cloud services and their CVEs instances within the knowledge graph, which enhances the understanding of the impact of CVEs of the sub-components of a particular cloud service on the end user of the cloud service.

\subsection{Apache Jena tools}
Apache Jena is a Java open-source framework used for building semantic web and linked data applications. It provides a set of libraries for RDF data, SPARQL queries, and semantic web applications. We implemented a parser tool using the Java programming language and Jena library \cite{ApacheJe10:online}. This tool automates the process of mapping data on cloud service components based on our ontology, thereby facilitating the identification and assessment of vulnerabilities in a more efficient and scalable manner.

\subsection{Use cases based on our ontology }

\begin{itemize}
    \item \textbf{ First use case : } we generate the knowledge graph as shown in figure \ref{fig:ServiceMode2CVE} to retrieve all linked CVEs on an instance of Platform as a service instance. 

    \begin{figure}[ht]
    \centering
    \includegraphics[scale=0.35]{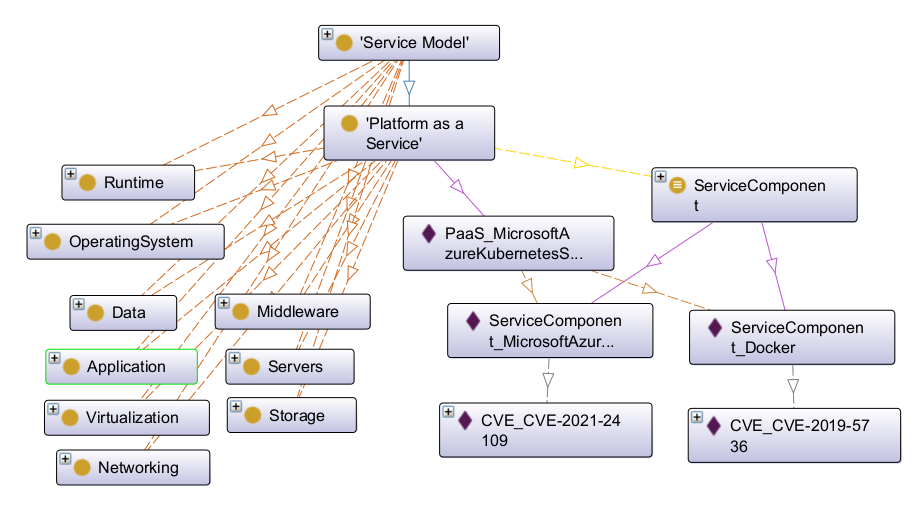}
    \caption{Display the CVEs for offered cloud services}
    \label{fig:ServiceMode2CVE}
    \end{figure}

\item \textbf{Second use case :} the knowledge graph generated to link  vulnerabilities into their CWE  as shown in figure \ref{fig:CWE_NG}.

\begin{figure}[ht]
    \centering
    \includegraphics[scale=0.43]{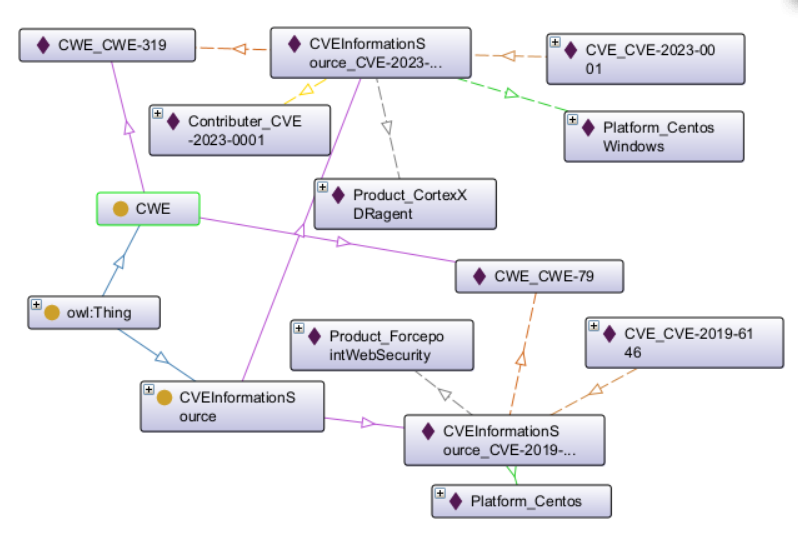}
    \caption{Vulnerabilities linked into their CWE}
    \label{fig:CWE_NG}
\end{figure}

\end{itemize}

\section{Discussion and Conclusions}
\label{sec:conclusion}

The description of vulnerabilities is important in cloud computing services, as it enables the end user to understand the weaknesses that the services may encounter and the appropriate mitigation methods. While existing sources of knowledge provide information at a highly abstract level, this study aims to delve deeper into the service stack to provide valuable insights at a lower level, focusing on service components. To achieve this, an ontology is proposed to describe and link vulnerabilities with comprehensive knowledge. Building upon previous efforts, we proposed an ontology that can be associated with each component in the service stack, encompassing all potential weaknesses. The proposed ontology consists of 18 classes and 17 object properties, providing a more detailed and comprehensive representation of vulnerabilities in cloud-computing services. 
In addition, we defined 26 data properties to further enhance the relationships and information associated with critical weaknesses related to specific components in cloud services. These data properties allow for a more detailed and specific mapping between vulnerabilities and their corresponding information, enabling a more comprehensive understanding of the security implication landscape on the services offered by cloud providers. Consequently, our proposed ontology not only provides a comprehensive representation of vulnerabilities in the offered service but also establishes connections between these vulnerabilities and the corresponding service components. By linking the weaknesses to specific subcomponents, our ontology enhances the understanding of how vulnerabilities can potentially impact different parts of the cloud-computing services offered. This enriched knowledge enables more targeted risk assessments, effective mitigation strategies, and improved overall security in cloud-based environments.

\printbibliography
\end{document}